\documentclass[pra,aps,twocolumn,groupedaddress,superscriptaddress,floatfix]{revtex4}
\usepackage{epsfig}
\usepackage{amsmath}
\allowdisplaybreaks[1]
\newcommand{\be}{\begin{equation}}
\newcommand{\ee}{\end{equation}}
\newcommand{\bea}{\begin{eqnarray}}
\newcommand{\eea}{\end{eqnarray}}
\newcommand{\ket}[1]{\left|#1\right\rangle}
\newcommand{\bra}[1]{\left\langle #1\right|}

\newcommand{\bc}{\begin{center}}
\newcommand{\ec}{\end{center}}

\renewcommand{\(}{\left(}
\renewcommand{\)}{\right)}
\renewcommand{\[}{\left[}
\renewcommand{\]}{\right]}
\newcommand{\forget}[1]{}
\newcommand{\half}{\dfrac{1}{2}}

\begin{document}
\title{Subwavelength atom localization via amplitude and phase control of
the absorption spectrum}
\author{Mostafa Sahrai}
\affiliation{Department of Physics, Tabriz University, Tabriz,
Iran}
\author{Habib Tajalli}
\affiliation{Department of Physics, Tabriz University, Tabriz,
Iran}
\author{Kishore T. Kapale}
\email{Kishor.T.Kapale@jpl.nasa.gov}
\affiliation{Quantum Computing Technologies Group, Jet Propulsion Laboratory,
California Institute of Technology, Mail Stop 126-347, 4800 Oak Grove Drive, 
Pasadena, California 91109-8099}

\author{M. Suhail Zubairy}
\affiliation{Institute for Quantum Studies and Department of
Physics, Texas A\&M University, College Station, TX 77843-4242}

\begin{abstract}
We propose a scheme for subwavelength localization of an atom conditioned upon 
the absorption of a weak probe field at a particular frequency. 
Manipulating atom-field interaction on a certain transition
by applying drive fields on nearby coupled transitions leads to interesting 
effects in the absorption spectrum of the weak probe field.
We exploit this fact and employ a four-level system with three driving fields
and a weak probe field, where one of the drive fields is a standing-wave 
field of a cavity. We show that the position of an atom along this standing wave is determined when probe field absorption is measured. 
We find that absorption of the weak probe field at a certain frequency
leads to subwavelength localization of the atom in either of the two
half-wavelength regions of the cavity field by appropriate choice of the system
parameters. We term this result as sub-half-wavelength localization to contrast it with the usual atom localization result of four peaks spread over one wavelength of the standing wave. We observe two localization peaks in either of the two half-wavelength regions along the cavity axis.
\end{abstract}
\maketitle

\section{Introduction}
Precision measurement of the position of an atom passing through a
standing-wave field has attracted considerable attention in recent years.
Several schemes have been proposed for the localization
of an atom using optical methods~\cite{welch9193}. High-resolution position measurements of the atom with optical
techniques are of considerable interest from both a theoretical and
an experimental point of view. Interest in the area 
is largely due to its applications to many areas of optical manipulations of atomic degrees of freedom, 
such as laser cooling~\cite{ChuMetcalf},
Bose-Einstein condensation~\cite{Collins96}, and atom
lithography~\cite{lithography} and measurement of the center-of-mass wavefunction of moving atoms~\cite{KapaleWavefunction}.

It is well known that optical methods provide better
spatial resolution in position measurements of atoms. For
example, in the optical virtual slits scheme the atom interacts
with a standing-wave field and  imparts a phase shift to the field. Measurement of this phase
shift then gives the position information of the atom~\cite{WallsZollerWilkens}.
Another related idea based on phase quadrature measurement is
considered in Ref.~\cite{Walls95}. Kunze {\it et al.}~\cite{Kunze97} demonstrated how 
the entanglement between the atomic position and its internal state
allows one to localize the atom without directly affecting its spatial 
wave function. It is shown that, by using Ramsey
interferometry, the use of a coherent-state cavity field is better
than the classical field to get a higher resolution in position
information of the atom~\cite{Kien97}. 
Resonance imaging methods have been used in experimental studies of the precision position measurement of the moving atoms~\cite{welch9193,Thomas90,Bigelow97}.

More recently, atom-localization
methods based on the detection of the spontaneously emitted photon
during the interaction of an atom with the classical standing-wave
field are considered~\cite{Zoller96,Herkomer97,QamarPRA2000,QamarOC2000}. 
We consider some of these proposals in detail to contrast them with the current proposal.
Qamar {\it et al.}~\cite{QamarPRA2000} suggested a simple scheme for  localization of an atom  by using a simple two-level system interacting with the classical standing-wave field. They showed that the frequency of the spontaneously emitted photon
carries information about the position of the atom. In another related
study, they used a three-level atom, where the upper two levels are
driven by a classical standing-wave field and spontaneously
emitted photon measurement from the upper level to lower level
gives information about the atomic position~\cite{QamarOC2000}.  It has been shown that coherent control of spontaneous emission in multilevel system gives line-narrowing and
even spontaneous emission quenching~\cite{CohContrlSE}. 
By using three drive fields, the phase and
amplitude control of the driving field on the spontaneous emission spectrum
in a four-level has been investigated by Ghafoor {\it et al.}~\cite{GhafoorPRA2000}.  This scheme was utilized further for localization of an atom during its motion on the
classical standing-wave field~\cite{GhafoorPRA2002}.  Thus observing a spontaneously emitted photon can lead to atom localization in a variety of systems.

It is, however,  important to note that from an 
experimental point of view, observation of a spontaneous emission spectrum is very
tricky and difficult. In this context, another scheme based on a three-level
$\Lambda$-type system interacting with two fields---a probe
laser field and a classical standing-wave coupling field---has been used
for atom localization by Paspalakis and Knight~\cite{Knight2001}. 
They observe that in the case of a weak  probe field, measurement of the population in the upper level
leads to subwavelength localization of the atom during its motion in the standing wave.  Thus in essence this scheme uses absorption of a probe field for atom localization.

In this article, we describe another method for the localization of an
atom in a standing-wave field based on electromagnetically
induced transparency~\cite{EIT}. We consider a four-level system driven by two
driving fields and a classical standing-wave field. A similar
scheme was recently considered by us~\cite{SahraiGroupVel} to demonstrate phase and
amplitude control of the group velocity of a weak probe field. The scheme is
very similar to the one used by Ghafoor {\it et al.} for phase and amplitude
control of the spontaneous emission~\cite{GhafoorPRA2000} and  atom 
localization~\cite{GhafoorPRA2002}.
However, their consideration was based on monitoring the spontaneous emission
properties of this scheme. In the present study
we consider the absorption properties of the weak probe field by determining the susceptibility of the system 
at the probe frequency. We show that the probe absorption observed at appropriate
frequencies localizes the atom in the classical standing-wave field. 
We also investigate the effects of the amplitude and phase
of the driving fields on the precision of localizing an atom. A novel feature observed by us as compared to the absorption based scheme of Paspalakis and Knight~\cite{Knight2001} is as follows: They observe four localization peaks in one wavelength of the standing-wave field---i.e., two each in the two half-wavelength regions. However, we obtain a more precise localization in the sense that for a certain choice of parameters we can confine the atom to one of the half-wavelength regions. We obtain two localization peaks that occur in one of the half-wavelength regions. We term this new domain of subwavelength localization as sub-half-wavelength localization. We would like to point out that this sub-half-wavelength localization has  already been proposed by Ghafoor {\it et al.}~\cite{GhafoorPRA2002}, in the context of monitoring spontaneous emission spectrum. However, they have not used the term sub-half-wavelength localization.

This article is organized as follows. In Sec.~\ref{Sec:Model} we introduce
our model and  give the basic equations and their solution to determine the susceptibility. In Sec.~\ref{Sec:Results} we study the behavior of the susceptibility
along the normalized position coordinate along the standing wave for a variety of system parameters. This gives us conditions on the controllable parameters to attain localization of the atom as it passes through the standing-wave optical field in the cavity. We give an analytical expression for the appropriate detuning parameters to obtain sub-half-wavelength localization. Finally we present our conclusions in Sec.~\ref{Sec:Conclusion}.

\section{Model and Equations}
\label{Sec:Model}
\begin{figure}[ht]
\includegraphics[scale=0.7]{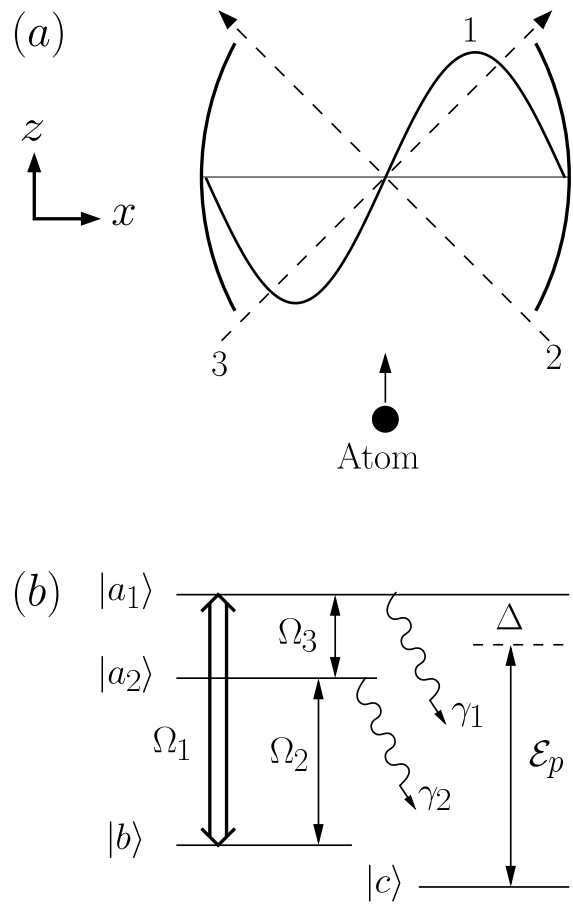}
\caption{\label{Fig:Scheme}The Model: $(a)$ The cavity supports the standing
wave field (1) corresponding to Rabi frequency $\Omega_1$. Two other fields
(2, 3) are applied at an angle as shown. The atom enters the cavity along the $z$
axis and interacts with the three drive fields. 
The whole process takes place in the $x$-$z$ plane. $(b)$ The energy level structure
of the atom. Probe field, denoted by $\mathcal{E}_p$, is detuned by an amount
$\Delta$ from the $\ket{a_1}$-$\ket{c}$ transition. The fields (2, 3) shown in the $(a)$
part of the figure correspond to the fields with Rabi frequencies $\Omega_2$ and
$\Omega_3$, respectively. The decay rates from the upper levels $\ket{a_1}$ and
$\ket{a_2}$ are taken to be $\gamma_1$ and $\gamma_2$, respectively.}
\end{figure}

The schematics of the proposed scheme are shown in Fig.~\ref{Fig:Scheme}. We
consider an atom, moving in the $z$ direction, as it passes through a
classical standing-wave field of a cavity. The cavity is taken to be aligned 
along the $x$ axis. The internal energy level structure of the atom is shown in 
Fig.~\ref{Fig:Scheme}(b). 
The radiative decay rates from the levels $\ket{a_1}$ and
 $\ket{a_2}$ to level $\ket{c}$ are taken to be $\gamma_1$ and $\gamma_2$.
The upper level $\ket{a_1}$ is coupled to the level
$\ket{a_2}$, and further the level $\ket{a_2}$ is coupled to level
$\ket{b}$ via classical fields with
Rabi frequencies $\Omega_3$ and $\Omega_2$, respectively. 
In addition, the upper level $\ket{a_1}$ is 
coupled to level
$\ket{b}$ via a classical standing-wave field of frequency $\nu$
and phase $\varphi$, having Rabi frequency $\Omega_1$. It should be noted that the Rabi frequency of the standing wave is  position dependent and is taken to be $\Omega_1(x)=\Omega_1 \sin \kappa x$. Here $\Omega_1(x)$ is defined to include the  position dependence and $\kappa$ is the wave vector of the standing-wave
field, defined as $\kappa =2 \pi/{\lambda}$, where $\lambda$ is the wavelength of the standing-wave field of the cavity.
We assume that the atom is initially in the state $\ket{c}$  and
interacts with a weak probe field that is near resonant with
the $\ket{c}\rightarrow\ket{a_1}$ transition. The detuning of the probe on this transition is taken to be $\Delta$. We  assume that the
center-of-mass position of the atom is nearly constant along the
direction of the standing wave. Therefore we apply the 
Raman-Nath
approximation and neglect the kinetic part of the atom from the
Hamiltonian~\cite{Meystre}. Under these circumstances, the Hamiltonian 
of the system in the
rotating-wave approximation can be written as
\begin{equation}
\label{eq:hamiltonian1} \mathcal{H} = \mathcal{H}_0 +
\mathcal{H}_I,
\end{equation}
where
\begin{equation}
\label{eq:hamiltonian2}\mathcal{H}_0 = \hbar
\omega_{a_1}\ket{a_1}\bra{a_1} + \hbar
\omega_{a_2}\ket{a_2}\bra{a_2} + \hbar\omega_b\ket{b}\bra{b} +
\hbar\omega_c\ket{c}\bra{c}
\end{equation}
and
\begin{multline}
\label{eq:hamiltonian3}\mathcal{H}_I=-\frac{\hbar}{2} \left[
\Omega_1 e^{- i \nu_1 t} \sin{\kappa x}\, \ket{a_1}\bra{b}\right.
\\
+ \Omega_2 e^{i k x \cos \theta_2} e^{- i
\nu_2 t}\ket{a_2}\bra{b}  \\ 
+ \left.\Omega_3 e^{i k x \cos{\theta_3}}  e^{- i \nu_3
t}\ket{a_1}\bra{a_2}
+\frac{\mathcal{E}_p\wp_{a_1 c}}{\hbar} e^{- i
\nu_p t}\ket{a_1}\bra{c}\right]+ \mbox{H.c.}
\end{multline}
Here $\omega_i$ are the frequencies of the states
$\ket{i}$ and $\nu_i$ are the frequencies of the 
optical fields, and  $\theta_{2}$ and  $\theta_{3}$ are the angles made by the propagation direction of the fields $\Omega_{2}$ and $\Omega_{3}$ with respect to the $x$ axis, respectively. The
subscript $p$ stands for the quantities corresponding to the probe
field; i.e., ${\mathcal{E}_p}$ and $\nu_p$ are the amplitude and 
frequency of the probe field. Also $\wp_{a_1 c}$ is the 
dipole matrix element of the $\ket{c}\rightarrow
\ket{a_1}$ transition. For simplicity we assume that
the Rabi frequencies $\Omega_1$ and $\Omega_2$ are 
real and $\Omega_3$ is complex---i.e., $\Omega_3=\Omega_3 e^{- i \varphi }$. This choice of imparting a carrying phase to field 3 is only for the convenience of calculations.  As will become clear later, only the relative phase of the three fields is important and absolute phases do not matter.
The dynamics of the system is
 described using density matrix approach as
\begin{equation}
\dot{\rho} = - \frac{i}{\hbar} [H,\rho]  - \frac{1}{2}
\{\Gamma,\rho\},
\end{equation}
where $\{\Gamma,\rho\} = \Gamma\rho + \rho \Gamma$. Here the decay
rate is incorporated into the equation by a relaxation matrix
$\Gamma$, which is defined by the equation $\langle n| \Gamma |
m \rangle = \gamma_n \delta_{nm}$. The detailed calculations of
these equations are given in the Appendix. 

Our goal is to obtain information about the atomic position from the susceptibility of 
the system at the probe frequency.  Therefore we need to determine the steady state value of the off-diagonal density matrix element  $\rho_{a_1 c}$. After necessary algebraic calculations and moving to appropriate rotating frames, we obtain a set of density
matrix equations [see Eq.~(\ref{Eq:completeset})].  To determine $\rho_{a_1 c}$  we only need following equations:
\begin{align}
\dot{\tilde{\rho}}_{a_1 c} &= - \bigl[i (\omega_{a_1 c}-\nu_p) +
\half\gamma_1\bigr]\tilde{\rho}_{a_1 c} + \frac{i}{2} \Omega_3 e^{- i \varphi }  e^{ i k x
\cos \theta_3 } \tilde{\rho}_{a_2 c} \nonumber \\ 
&\qquad+  \frac{i}{2} \Omega_1
\sin \kappa x \tilde{\rho}_{b c} - i
\frac{\mathcal{E}_p \wp_{a_1 c} }{2 \hbar} (\tilde{\rho}_{a_1 a_1}
- \tilde{\rho}_{cc}), 
\nonumber \\ 
\dot{\tilde{\rho}}_{a_2 c} &= -\bigl[i (\omega_{a_2 c} - (\nu_p- \nu_3)) + \half\gamma_2\bigr] \tilde{\rho}_{a_2 c} 
\nonumber \\
&\qquad+ 
  \frac{i}{2} \Omega_2 e^{ i k x \cos
\theta_2 }\tilde{\rho}_{b c}  +
\frac{i}{2} \Omega_3 e^{ i \varphi }e^{- i k x \cos \theta_3 }
\tilde{\rho}_{a_1 c}
\nonumber \\ 
&\qquad \qquad- i \frac{\mathcal{E}_p \wp_{a_1 c}}{2 \hbar} \tilde{\rho}_{a_2 a_1}, 
\nonumber \\
\dot{\tilde{\rho}}_{b c} &= - [i ( \omega_{b c} + \nu_1 -\nu_p ) +
\gamma_{b c}]\tilde{\rho}_{b c} + \frac{i}{2} \Omega_1 \sin \kappa x \tilde{\rho}_{a_1 c} 
\nonumber \\
&\qquad+ \frac{i}{2}\Omega_2 e^{- i k x \cos \theta_2 } \tilde{\rho}_{a_2 c} - i
\frac{\mathcal{E}_p \wp_{a_1 c}}{2\hbar} \tilde{\rho}_{b a_1}.
\label{eq:tilderhodot}
\end{align}

As we know, the dispersion and absorption are related to the
susceptibility of the system and are determined by $\rho_{a_1 c}$.
We take the probe field to be weak and  calculate the
polarization of the system to lowest order in $\mathcal{E}_p$.
We keep all the terms of the driving fields but keep
only linear terms in the probe field. The atom is 
initially in the ground state $\ket{c}$; therefore, we use
\begin{equation}
\tilde{\rho}_{cc}^{(0)} =1, \quad \tilde\rho_{b a_1}^{(0)}=0,
\quad \tilde\rho_{a_2 a_1 }^{(0)} = 0, \quad \tilde\rho_{a_1
a_1}^{(0)} = 0.
\label{Eq:initial}
\end{equation}
Equations~(\ref{eq:tilderhodot}) can then be simplified considerably to obtain
\begin{align}
\dot{\tilde{\rho}}_{a_1 c} &= - \bigl(i \Delta +
\half\gamma_1\bigr)\tilde{\rho}_{a_1 c} + \frac{i}{2} \Omega_3\, e^{- i \varphi } e^{ i k x
\cos \theta_3 } \tilde{\rho}_{a_2 c} 
\nonumber \\
&\qquad+  \frac{i}{2} \Omega_1 \sin{\kappa x} \,\tilde{\rho}_{b c} + i
\frac{\mathcal{E}_p \wp_{a_1 c} }{2 \hbar} , 
\nonumber \\
\dot{\tilde{\rho}}_{a_2 c} &= - \bigl(i \Delta + \half\gamma_2\bigr)\tilde{\rho}_{a_2 c} 
+ \frac{i}{2} \Omega_3\, e^{i \varphi } e^{- i k x \cos \theta_3 } \tilde{\rho}_{a_1 c} 
\nonumber \\
&\qquad+  \frac{i}{2} \Omega_2 e^{i k x \cos \theta_2 }\tilde{\rho}_{b c}, \nonumber \\
\dot{\tilde{\rho}}_{b c} &= - i\, \Delta\, \tilde{\rho}_{b c} +
\frac{i}{2} \Omega_1  \sin{\kappa x} \,\tilde{\rho}_{a_1 c}
\nonumber \\
&\qquad +  \frac{i}{2} \Omega_2 e^{- i k x \cos
\theta_2 } \tilde{\rho}_{a_2 c}. \label{eq:tilderhodot2}
\end{align}
Here we have introduced the detuning of the probe field and the 
frequency difference between levels $\ket{a_1}$ and $\ket{c}$,
\begin{equation}
\Delta = \omega_{a_1c} - \nu_p = \omega_{a_2c}+ \nu_3 - \nu_p =
\omega_{bc} + \nu_1 - \nu_p.
\label{Eq:Delta}
\end{equation}
We have also assumed $\gamma_{b c}=0$.

This set of equations can be solved
analytically. Following the treatment discussed in the  Appendix, the off-diagonal
density-matrix element corresponding to  the probe transition is obtained as
\begin{equation}
{\rho}_{a_1 c} = \tilde{\rho}_{a_1 c} e^{- i \nu_p t} =
\frac{1}{Y \hbar} (\Omega_2^2 - 4 \Delta^2 +  2 i \gamma_2 \Delta)
\mathcal{E}_p \wp_{a_1 c} e^{- i \nu_p t},
\label{eq:tilderhoa1c}
\end{equation}
where we have assumed $\theta_3=\pi/ 4$, 
$\theta_2=\pi/2+\pi/ 4$,
and $Y$ is given by
\begin{align}
Y = A + i B,
\end{align}
with
\begin{align}
A &= - 8 \Delta^3 +  2 \Delta (\Omega_1^2 \sin^2 \kappa x + \Omega_2^2 +
\Omega_3^2  )
\nonumber \\
&\qquad + 2 \gamma_1 \gamma_2 \Delta +
\Omega_1 \Omega_2 \Omega_3 (e^{i \varphi} + e^{-i \varphi}) \sin
\kappa x , \nonumber \\ B &= 4 \Delta^2 (\gamma_1 +\gamma_2) - 
\gamma_1 \Omega_2^2 - \gamma_2 \Omega_1^2 \sin^2 \kappa x.
\end{align}

The susceptibility at the probe frequency can be written as
\begin{equation}
\chi =\frac{2 N \wp_{a_1c} \rho_{a_1c} }{\epsilon_0 \mathcal{E}_p }e^{i \nu_p t} 
      =\frac{2 N |\wp_{a_1c}|^2}{\epsilon_0 } \frac{ (\Omega_2^2 - 4 \Delta^2 + 2 i  \gamma_2 \Delta) }{Y \hbar},
\end{equation}
where $N$ is the atom number density in the medium. The real and imaginary
parts of susceptibility are given as
\begin{align}
\chi' & = \frac{2 N |\wp_{a_1 c}|^2 }{ \epsilon_0 \hbar Z} \{
(\Omega_2^2 - 4 \Delta^2) A + 2 \gamma_2 \Delta B\},\\ \chi'' & =
\frac{2 N |\wp_{a_1 c}|^2 }{ \epsilon_0 \hbar Z} \{ 2  \gamma_2
\Delta A - (\Omega_2^2 - 4 \Delta^2) B)\},
\label{Eq:Chiprime}
\end{align}
where $Z = Y Y^*$ and $\chi=\chi'+ i \chi''$. It is imperative to point out that the phase enters the susceptibility expression only through the quantities $A$ and $Y$. Even the phase dependence of $Y$ is only through $A$. Moreover, we observe that the 
phase-dependent term in $A$ is $\Omega_1 \Omega_2 \Omega_3 (e^{i \varphi} + e^{-i \varphi}) \sin \kappa x $. Thus the phase factor could very well have come from either of the three driving fields. As pointed out earlier, if all the fields had phase dependence, only the collective phase would be important and no individual phase-dependent terms would occur. This is because  the Rabi frequencies $\Omega_{i}$ in all the other terms appear through $\Omega_{i}^{2}$, which is $|\Omega_{i}|^{2}$ for a complex Rabi frequency $\Omega_{i}=|\Omega_{i}|e^{i \phi_{i}}$. The collective phase can be easily determined to be $\phi=\phi_{2}+\phi_{3}-\phi_{1}$ by repeating the susceptibility calculation. Here $\phi_{i}$ is  the phase of the complex Rabi frequency $\Omega_{i}$ of the $i$th driving field.

In the next section we consider the imaginary part of the susceptibility
$\chi''$ in detail and obtain various conditions for subwavelength localization of the atom.

\section{Results and Discussion}
\label{Sec:Results}

We study  expression~(\ref{Eq:Chiprime}) for the imaginary part of the susceptibility on the probe transition in greater detail in the following 
discussion. It is clear that $\chi''$---i.e., the probe absorption---depends 
on the controllable parameters of the system  like probe field detuning and
amplitudes and phases of the driving fields. 

Noting the dependence of $\chi''$ on $\sin{\kappa x}$, it  is, in principle, possible to  obtain information
about the $x$ position of the atom as it passes through the cavity 
by measuring the probe absorption. Nevertheless, for precise localization of the
atom the susceptibility should show maxima or peaks along the
$x$ coordinate. We obtain the conditions for the presence of peaks in $\chi''$ in the discussion to follow. In the case of $\gamma_{2}=0$---i.e., the level $\ket{a_{2}}$ is  metastable---Eq.~(\ref{Eq:Chiprime}) can be simplified as follows:
\begin{widetext}
\begin{align}
\chi'' &= \frac{2 N |\wp_{a_1c}|^2}{\hbar \epsilon_0} 
\frac{\gamma_1 (\Omega_2^2 - 4 \Delta^2)^2}
{\gamma_1^2(\Omega_2^2 -4\Delta^2 )^2 + [ 8 \Delta^3 
- 2\,\Delta \,(\Omega_1^2\,\sin^2{\kappa x} + \Omega_2^2 + \Omega_3^2)
-  2\,\Omega_1 \Omega_2 \Omega_3\, \cos\varphi\, \sin{\kappa x}]^2}
\nonumber \\
&=\frac{2 N |\wp_{a_1c}|^2}{\hbar \epsilon_0} 
\frac{\gamma_1 (\Omega_2^2 - 4 \Delta^2)^2}
{\gamma_1^2(\Omega_2^2-4\Delta^2 )^2 
+ 4 \Delta^2 \Omega_1^4\, (\sin{\kappa x}-R_1)^2(\sin{\kappa x}-R_2)^2},
\label{Eq:chiroot}
\end{align}
where
\begin{equation}
R_{1,2}=\frac{1}{2
\Delta \Omega_1}\left\{ {-\Omega_2 \Omega_3 \cos\varphi}
\pm \sqrt{\Omega_2^2\, \Omega_3^2\, \cos^2{\varphi} - 4 \Delta^2 [ (
\Omega_2^2 + \Omega_3^2 )- 4 \Delta^2 ]}\right\}\,.
\end{equation}
\end{widetext}
From Eq.~\eqref{Eq:chiroot} it is clear that peaks will occur in  $\chi''$ at $x$ positions satisfying 
$\sin{\kappa x } = R_{1,2}$. In other words, the probe absorption peaks
at the spatial position defined by
\begin{align}
\kappa x& = \sin^{-1} \Biggl[ \left.\frac{1}{2
\Delta \Omega_1}\right\{ {-\Omega_2 \Omega_3 \cos\varphi}
\nonumber \\
&\quad\pm \left.\sqrt{
\Omega_2^2\, \Omega_3^2\, \cos^2{\varphi} - 4 \Delta^2 [ (
\Omega_2^2 + \Omega_3^2 )- 4 \Delta^2 ]}\right\}\Biggr]
\nonumber \\
&\qquad\pm n \pi,
\end{align}
where $n$ is an integer. This leads to localization of atoms conditioned on probe absorption at a particular frequency corresponding to the value of $\Delta$. 

\begin{figure*}[ht]
\includegraphics[scale=0.44]{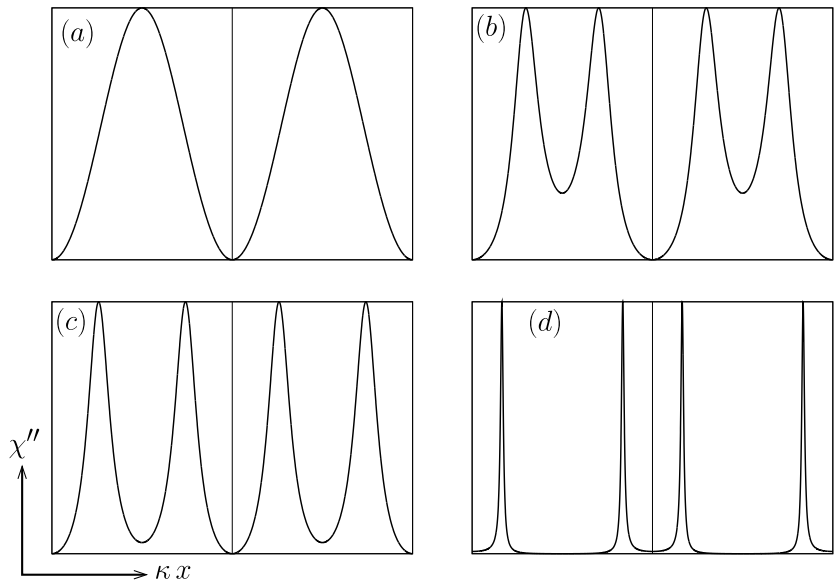}
\caption{\label{Fig:plots1} Dependence of the probe absorption on drive-field  Rabi frequencies and detuning: plot of the imaginary part of the susceptibility in
arbitrary units vs the dimensionless $x$ coordinate $\kappa x$ along the standing
wave in the cavity. $\kappa x$ runs from the values $-\pi$ on the extreme left
to $\pi$ to the extreme right in each box. A vertical line is drawn at $\kappa x
= 0$ to guide the eye. The common parameters are 
$\Omega_2= \Omega_3=\Omega=\gamma_1$, $\gamma_2=0$, $\Omega_1 = 3\gamma_1$, and  $\varphi=\pi/2$ unless specified otherwise.
$(a)$ $\Delta= 5\gamma_1$,
$(b)$ $\Delta=   1.4 \gamma_1$,
$(c)$ $\Delta=   1.3\gamma_1$,
and  $(d)$ $\Omega_1 = 20\gamma_1$, $\Delta=5\gamma_1$.
Notice that with the increase in the strength of the standing-wave field the peaks get sharper. Although the detuning is the same for $(a)$ and $(b)$, the number of peaks  is different as the cavity-field strength is different.}
\end{figure*}

We show the dependence of $\chi''$ or probe absorption in arbitrary units versus the
dimensionless $x-$coordinate in Fig.~\ref{Fig:plots1}. We show how the number of
peaks, their positions, and widths vary  as the probe-field detuning, and driving field Rabi frequencies and the relative phase carried by the standing-wave field are changed. 
As seen from $(a)$-$(c)$ in Fig.~\ref{Fig:plots1},  the number of peaks is dependent on the detuning.  As the detuning is  decreased the peaks separate and we obtain four peaks in $(b)$ and $(c)$ compared to only two in $(a)$. In $(d)$ we show that, for a larger strength of the standing-wave cavity field, the peaks become sharp, leading to localization of the atom at one of the four possible positions. It can be noted that such localization is  conditioned on the measurement of the absorption of the probe field at a frequency corresponding to the chosen value of the detuning. 

The positions of the probe absorption maxima are strongly dependent on the 
probe-field frequency through its detuning $\Delta$ defined in Eq.~(\ref{Eq:Delta}).   To clarify this point further we consider the relation $\sin \kappa x = R_{1,2}$ for various values of the phase parameter $\varphi$ and solve for the detuning. This gives the values of the detunings as a function of $\kappa x$ to obtain probe absorption peaks, provided all other parameters are fixed.  Taking $\Omega_2 = \Omega_3 = \Omega$ for simplicity we obtain
\begin{widetext}
\begin{align}
\label{Eq:Detuning1}
\Delta & =\frac{1}{4}\,\[\Omega_1 \sin \kappa x \pm \sqrt{\Omega_1^2 \sin^2 \kappa x + 8 \Omega^2}\]= \delta^{(0)}_{1,2} \quad \text{ or }\quad \Delta = -\frac{1}{2}\,{\Omega_1 \sin \kappa x} = \delta^{(0)}_{3}\quad\text{ for } \varphi = 0,  \\
\label{Eq:Detuning2}
\Delta &= \pm \frac{1}{2}\, \sqrt{\Omega_1^2 \sin^2 \kappa x + 2 \Omega^2}=\pm\,\delta^{(\pi/2)}_1 \quad\text{ for }  \varphi = \pi/2,  \\
\label{Eq:Detuning3}
\Delta &=\frac{1}{4}\,[-\Omega_1 \sin \kappa x \pm \sqrt{\Omega_1^2 \sin^2 \kappa x + 8 \Omega^2}]=\delta^{(\pi)}_{1,2}\quad\text{ or }\quad \Delta=\frac{1}{2}\,{\Omega_1 \sin \kappa x}= \delta^{(\pi)}_{3}\quad\text{ for } \varphi = \pi\,.
\end{align}
\end{widetext}
From these relations it can be seen that there is  a strong phase dependence on the detuning values required to obtain  probe absorption maxima. 
Also for each phase value there are several detuning values that would give rise to probe absorption maxima at each position. Namely, for $\varphi=0$ and $\pi$ there are three different values given by $\delta^{(0)}_{1,2,3}$ and $\delta^{(\pi)}_{1,2,3}$, respectively. Similarly for $\varphi=\pi/2$  detuning can take either of the values $\pm\delta_{1}^{(\pi/2)}$. 

\begin{figure*}[ht]
\includegraphics[scale=0.44]{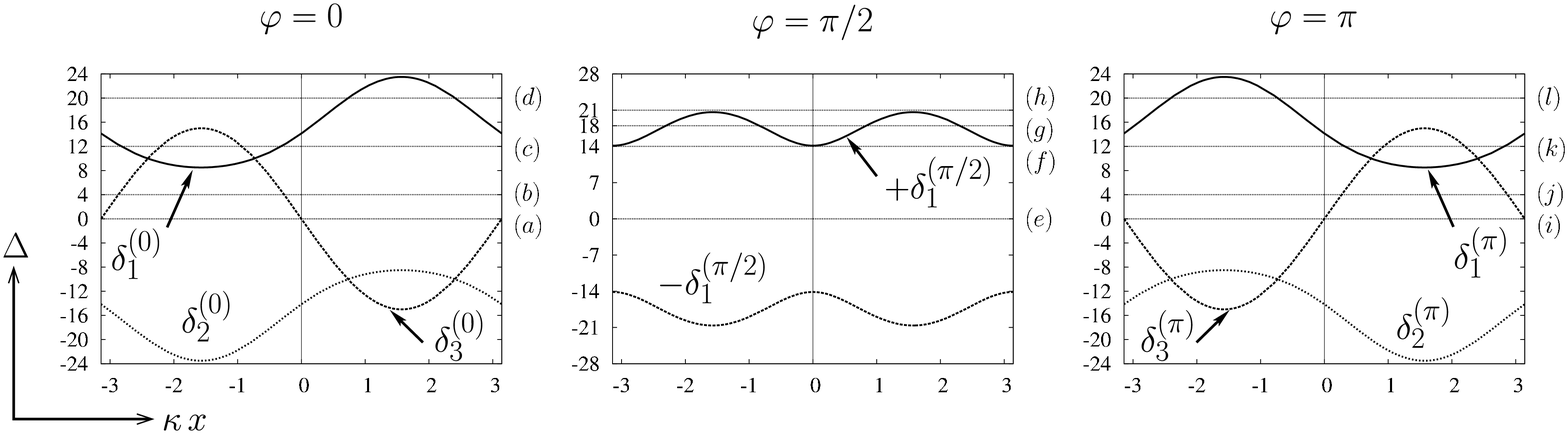}
\caption{\label{Fig:plots-detuning} Plot of the RHS of Eqs. \eqref{Eq:Detuning1}--\eqref{Eq:Detuning3} (in units of $\gamma_1$) vs the dimensionless $x$ coordinate $\kappa x$. Each curve gives the values of the detunings at given positions corresponding to peaks in the probe absorption. The common parameters are taken to be $\Omega_2= \Omega_3=\Omega=20\gamma_1$, $\gamma_2=0$, and $\Omega_1 = 30\gamma_1$. Note that the character of these plots would change for different 
drive-field parameters. Interesting values of the detunings chosen for further analysis are shown by horizontal lines in each graph. The number of places at which the horizontal line intersects with the plotted curves corresponds to the number of peaks observed in the probe absorption, and the corresponding $x$ coordinate  gives the positions of the peaks in the cavity standing-wave field.}
\end{figure*}

To study this dependence in depth we plot the right-hand side (RHS) of the Eqs.~\eqref{Eq:Detuning1}--\eqref{Eq:Detuning3} at different positions in one wavelength of the cavity standing-wave field in Fig.~\ref{Fig:plots-detuning}. These curves give the detuning value required to obtain peaks in the probe absorption at the corresponding position along the  cavity axis. However, in an experiment  a particular value for the detuning of the probe needs to be chosen to begin with.  Therefore, for a given fixed detuning, represented by horizontal lines in the plots, the maximum conditions would be satisfied only if the horizontal lines intersect with the curves obtained from the solution of Eqs.~\eqref{Eq:Detuning1}--\eqref{Eq:Detuning3}.  That is, there will be certain $x$ positions at which probe absorption maxima will occur. The curves obtained for $\varphi=0$ and $\varphi=\pi$ happen to be related to each other by the transformation $x\rightarrow -x$. Moreover, there exists another symmetry for the phase values $\varphi=0,\pi$. The curves obtained by changing the sign of the detuning are  the same as the original ones transformed according to $x\rightarrow -x$. Whereas, the curves for $\varphi=\pi/2$ are entirely different and the structure is independent of the sign of the detuning. Keeping these symmetries in mind we only consider positive values of the detunings, corresponding to a probe frequency less than the probe transition frequency, for the rest of the discussion. The results for negative values of the detunings can be similarly obtained.

\begin{figure*}[ht]
\includegraphics[scale=0.31]{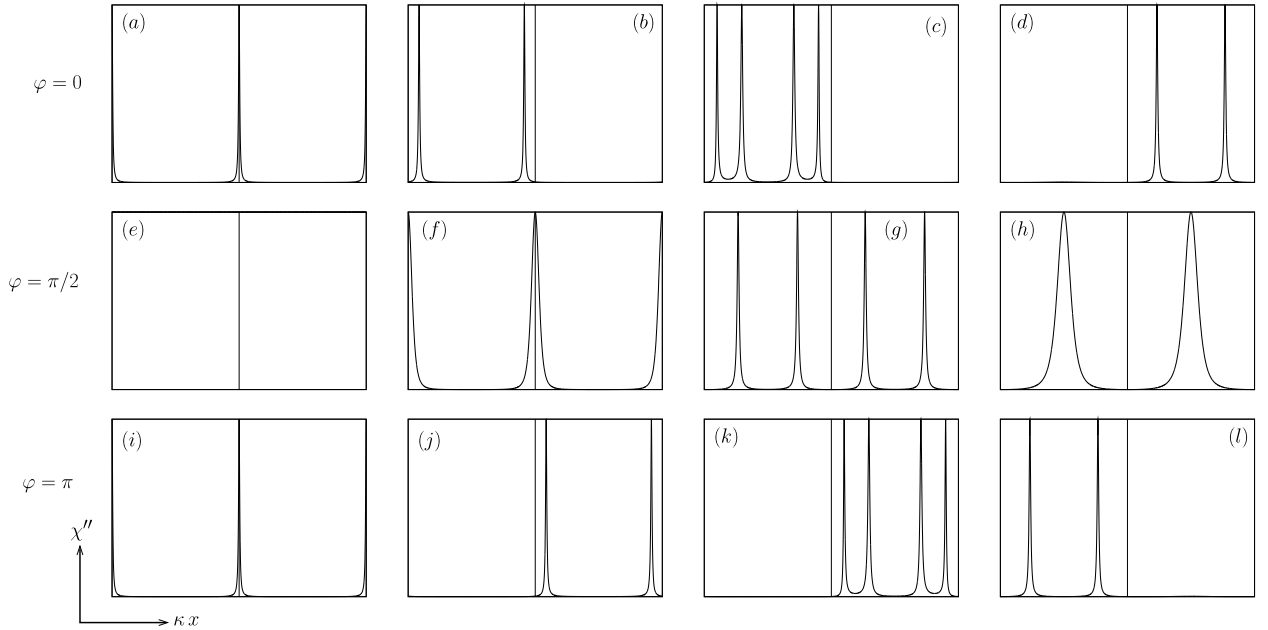}
\caption{\label{Fig:plots-phase} Phase dependence of the localization for several values of the detuning parameter. Plot of the imaginary part of the susceptibility in arbitrary units vs the dimensionless $x$ coordinate $\kappa x$. $\kappa x$ runs from the values $-\pi$ on the extreme left to $\pi$ to the extreme right in each box. A vertical line is drawn at $\kappa x = 0$ to guide the eye. The common parameters are the same as in Fig.~\ref{Fig:plots-detuning}. The labels $(a)$--$(l)$ correspond to the detuning values similarly labeled in Fig.~\ref{Fig:plots-detuning}. Notice sub-half-wavelength localization for boxes $(b)$, $(c)$, and $(d)$ for phase $\varphi=0$. Similarly, for $\varphi=\pi$ sub-half-wavelength localization exists as shown in boxes $(j)$, $(k)$, and $(l)$. The term sub-half-wavelength localization is coined for a special regime of atom localization where the localization peaks are confined to either the range $\kappa x = \{-\pi, 0\}$ or $\kappa x = \{ 0, \pi \}$. It can be noted that for $\varphi= \pi/2$ there is no sub-half-wavelength localization, although the number of peaks varies according to the detuning value chosen. The interesting results correspond to boxes $(d)$ and $(l)$ where there are only two peaks that are confined to a half-wavelength region on the cavity field. In box $(e)$ the probe absorption is the same at all positions in the cavity field represented by the dark line appearing at the top boundary of the box. }
\end{figure*}

It can be seen from the plots in Fig.~\ref{Fig:plots-detuning} that the number of intersecting points and their positions depend on the value of the detuning chosen as well as the relative phase $\varphi$ of the drive fields. The points of intersection correspond to the probe absorption maxima. In Fig.~\ref{Fig:plots-detuning} we have considered several values of the detunings for different phase parameters. From the figure the positions and number of peaks to be observed in the probe field absorption can be predicted. 
To understand this in detail we consider these special detuning values -labeled  $(a)$--$(l)$ in Fig.~\ref{Fig:plots-detuning}] and plot numerically evaluated $\chi''$ in Fig.~\ref{Fig:plots-phase}.

It is straightforward to make contact with the results presented in Figs.~\ref{Fig:plots-detuning} and \ref{Fig:plots-phase}. For each value of the detuning, represented by horizontal lines in Fig.~\ref{Fig:plots-detuning}, we have a box showing the corresponding probe absorption curve along the cavity standing-wave field. The predictions about the number of peaks and their position for each detuning and phase value made by observing Fig.~\ref{Fig:plots-detuning} can be verified by looking at the corresponding box in Fig.~\ref{Fig:plots-phase}. Thus, we observe that the number of peaks and their positions depend on the phase and detuning values as long as other parameters are kept fixed. Using Fig.~\ref{Fig:plots-phase} we can make several observations. 

We obtain an interesting regime of atom localization for $\varphi=0\text{ and }\pi$; the probe absorption peaks are situated in either $\kappa x = \{-\pi,0\}$ or  $\kappa x = \{0,\pi\}$ half-wavelength regions along the cavity field, provided the detunings are chosen properly [see Fig.~\ref{Fig:plots-phase}, boxes $(b)$, $(c)$, $(d)$, $(j)$, $(k)$, and $(l)$]. The more interesting regime corresponds to the existence of just two peaks as opposed to the usual four. We have coined the term sub-half-wavelength localization for such an interesting regime of localization. It can be seen that for $\varphi=\pi/2$, although the number of peaks and their widths vary as per the detuning value, there is no sub-half-wavelength localization possible. Neverthless, the usual atom localization regime [See Fig.~\ref{Fig:plots-phase}, box $(g)$] is still available. Another interesting observation for the parameters $\varphi=\pi/2, \Delta=0$ [See $(e)$] gives no peaks in the probe absorption. The probe absorption is uniform over the length of the cavity, although the cavity field strength is different at different points. As seen from Fig.~\ref{Fig:plots-detuning}, this value of the detuning does not intersect with the solutions of the maximum condition $\sin \kappa x = R_{1,2}$. For all the results represented in this figure we have chosen positive values for the probe detuning; however, the results can be similarly obtained for negative detunings as well and they show similar characteristics.

It is imperative to clarify the results obtained here. It can be seen that observing the peak in the absorption of the probe field of a particular frequency leads to localization of the atom. It means only if the atom passed through that position along the cavity field, one would obtain peak absorption for the probe.  Another atom passing through the cavity may not pass through the position corresponding to the absorption peak; in such a case, the probe field will not experience any absorption. This can be alternatively understood as the center-of-mass distribution of the atomic beam is not changed by the procedure considered here. It is only modulated by the coherent processes occurring in the system. Therefore, the localization of the atom is conditioned upon the observance of a peak in the absorption of  the probe at particular frequency.  However, the fact that a particular frequency is associated with a particular location along the standing wave, one can envision the probe field carrying multiple frequencies. Such a configuration can then lead to localization of atoms at different locations conditioned upon the peak absorption of the corresponding  frequency component of the probe.  Such a scheme can be applied for atom lithography for generating arbitrary one-dimensional patterns. 
One can extend these arguments for generating arbitrary two dimensional patterns as well. The advantage our model provides is that  the resolution is much larger compared to conventional subwavelength localization. We have succeeded in localization in the sub-half-wavelength domain. 

Thus, we have shown how to obtain sub-half-wavelength localization through monitoring the probe absorption at a particular frequency.  We note that the atom is to be prepared in its ground state to start with  [See Eq.~(\ref{Eq:initial})] as opposed to schemes based on the observation of the  spontaneous emission spectrum (e.g., Ref.~\cite{GhafoorPRA2002}), where the atom needs to be prepared in its excited state. Thus the preparation stage is considerably simplified in our model. Moreover, as we need monitoring the probe absorption as opposed to spontaneous emission as in 
Ref.~\cite{GhafoorPRA2002}, we have a distinct advantage to offer as the absorption measurements are straightforward to realize in an experiment compared to the measurement of the spontaneous emission spectrum. In the following section we summarize our conclusions.

\section{Conclusion}
\label{Sec:Conclusion}

We have presented a scheme for subwavelength localization  of a moving atom as it passes through  the standing-wave field of a cavity.  
This allows us to determine the position of the moving atoms with high precision as we show the possibility of confinement of the atom to a sub-half-wavelength region. Our 
scheme is based on the measurement of absorption of a weak probe field by the atom. Measurement of
the absorption  of a weak probe field of prechosen frequency localizes the
atom in real time. We have shown that the precision of the  position measurement of the atom  depends upon the amplitude and phase of
the classical standing-wave field. The amplitude of standing-wave driving field when increased leads to  line narrowing in the probe absorption, thus giving increased precision in the position measurement whereas the phase of the standing-wave driving
field has important role in reducing the number of localization
peaks from the usual four to two, leading to a new localization scheme which we call sub-half-wavelength localization.  Moreover, we show that the proper choice of probe frequency is very important in obtaining  sub-half-wavelength localization. 
As our method is based on the measurement of the probe absorption, it has two distinct advantages compared to similar methods  based on observation of the spontaneous emission spectrum. Absorption measurements are much easier to perform in a laboratory compared to monitoring of the spontaneous emission spectrum. Moreover, we do not require the atoms to be prepared in their excited states. We require that  they are prepared in their ground state, which is a fairly routine task in atomic physics experiments.  Thus the preparation stage is fairly straightforward. These advantages suggest an easy experimental  implementation of our scheme.

\acknowledgments
Part of this work was carried out (by K.T.K.)
at the Jet Propulsion Laboratory under 
a contract with the National Aeronautics and Space Administration (NASA). 
K.T.K. acknowledges support from the National Research Council and
NASA, Code Y.  M.S.Z. acknowledges  support of the Air
Force Office of Scientific Research,
DARPA-QuIST, TAMU
Telecommunication and Informatics Task Force (TITF) Initiative,
and the Office of
Naval Research. 

\appendix
\section*{Details of the susceptibility calculations}
\label{appendix-details}

Here we give details of the derivation of the density matrix
equation and their solution for the calculation of the susceptibility. 
The density matrix equations 
are given by Eq.(4), so the $(i,j)$th element of the density matrix satisfies the
equation
\begin{equation}
\dot{\rho_{ij}} = - \frac{i}{\hbar} \sum_{k}(H_{ik},\rho_{kj} -
\rho_{ik} H_{kj}) - \frac{1}{2} \sum_k (\Gamma_{ik}\rho_{kj} +
\rho_{ik}\Gamma_{kj}).
\label{Eq:appeqn1}
\end{equation}
Here the indices $i, j$, and $k$ run over $a_1$, $a_2$, $b$ and $c$ . The
decay matrix has only two nonzero elements given by $\Gamma_{a_1 c} =
\gamma_1$  and $\Gamma_{a_2 c} = \gamma_2$. By utilizing the
Hamiltonian given in Eqs.~(1)--(3) and Eq.~(\ref{Eq:appeqn1}), the
off-diagonal density matrix elements can be shown to satisfy the following set of equations:
\begin{align}
\dot{\rho}_{a_1a_2} &= -\Bigl(i\, \omega_{a_1a_2} +\half(\gamma_{1}+\gamma_{2})\Bigr)\rho_{a_1a_2} 
\nonumber \\
&\quad
-\frac{i}{2} \Omega_3\, e^{-i \varphi} e^{i k x \cos \theta_3} e^{-i \nu_3 t}(\rho_{a_1a_1} - \rho_{a_2a_2}) 
\nonumber \\
&\quad+ \frac{i}{2} \Omega_1  \sin{\kappa x}\, e^{-i \nu_1 t} \rho_{ba_2} 
\nonumber \\ 
&\quad - \frac{i}{2} \Omega_2\, e^{- i k x \cos \theta_2} e^{i
\nu_2 t} \rho_{a_1b} + \frac{i \mathcal{E}_p \wp_{a_1c}}{2\hbar}
e^{-i \nu_p t} \rho_{ca_2}, 
\nonumber \\ 
\dot{\rho}_{a_1b} &= - \Bigl( i\, \omega_{a_1b} +\half \gamma_{1} \Bigr) \rho_{a_1b} 
\nonumber \\
&\quad
- \frac{i}{2} \Omega_1 \sin{\kappa x}\, e^{-i \nu_1 t} (\rho_{a_1a_1} - \rho_{bb}) 
\nonumber \\
&\quad - \frac{i}{2}\Omega_2\, e^{i k x \cos \theta_2} e^{-i
\nu_2 t} \rho_{a_1a_2} 
\nonumber \\ 
& \quad+ \frac{i}{2}\Omega_3\,  e^{- i \varphi} e^{ i k x \cos \theta_3} e^{-i \nu_3 t} \rho_{a_2b} 
+ \frac{i \mathcal{E}_p \wp_{a_1c}}{2\hbar} e^{-i \nu_p t} \rho_{cb},
\nonumber \\ 
\dot{\rho}_{a_2b} &= - \Bigl( i\, \omega_{a_2b} +\half \gamma_{2} \Bigr)\rho_{a_2b} 
\nonumber \\
&\quad
-\frac{i}{2} \Omega_2\, e^{ i k x \cos \theta_2 } e^{-i \nu_2 t} (\rho_{a_2a_2} - \rho_{b b}) 
\nonumber \\
&\quad-\frac{i}{2}\Omega_1  \sin{\kappa x}\, e^{-i \nu_1 t} \rho_{a_2a_1} 
\nonumber \\ &
\quad + \frac{i}{2} \Omega_3\, e^{i \varphi} e^{-i k x \cos \theta_3} e^{i
\nu_3 t} \rho_{a_1b}, 
\nonumber \\ 
\dot{\rho}_{a_1c} &= -\Bigl(i\, \omega_{a_1c} + \half \gamma_1 \Bigr)\rho_{a_1c} 
+ \frac{i}{2} \Omega_3\, e^{- i \varphi} e^{i k x \cos \theta_3} e^{-i \nu_3 t} \rho_{a_2c} 
\nonumber \\
&\quad+ \frac{i}{2} \Omega_1 \sin{\kappa x} \,e^{-i \nu_1 t} \rho_{bc} 
\nonumber \\ 
& \quad - \frac{i \mathcal{E}_p \wp_{a_1c}}{2\hbar} e^{-i \nu_p t} (\rho_{a_1a_1}-\rho_{cc}),
\nonumber \\ 
\dot{\rho}_{a_2c} &= -\Bigl(i\, \omega_{a_2c} + \half \gamma_2 \Bigr)\rho_{a_2c} 
+ \frac{i}{2} \Omega_3\, e^{ i \varphi} e^{- i k x \cos \theta_3} e^{i \nu_3 t} \rho_{a_1c} 
\nonumber \\
&\quad+ \frac{i}{2}\Omega_2 e^{i k x \cos \theta_2} e^{-i \nu_2 t} \rho_{bc} 
\nonumber \\ &
\quad - \frac{i \mathcal{E}_p \wp_{a_1c}}{2\hbar} e^{-i \nu_p t}\rho_{a_2a_1}, \nonumber \\ 
\dot{\rho}_{bc} &= -\Bigl(i\,\omega_{bc}+\gamma_{bc} \Bigr)
\rho_{bc} + \dfrac{i}{2} \Omega_1 \sin{\kappa x} \,e^{i \nu_1 t} \rho_{a_1c}
\nonumber \\
&\quad+ \dfrac{i}{2} \Omega_2 e^{- i k x
\cos \theta_2}e^{i \nu_2 t} \rho_{a_2c} - \frac{i \mathcal{E}_p
\wp_{a_1c}}{2\hbar} e^{-i \nu_p t}\rho_{ba_1}.
\end{align}

The diagonal density matrix elements---i.e., the populations of the atomic energy
levels---can be determined in a similar manner. Here $\omega_{i k}$
corresponds to the energy difference between the levels $i$, $k$, and
$\nu_k$  is the frequency of the $k$th optical field. Next we
transform these equations to appropriate rotating frame
defined through
\begin{align}
\rho_{c a_2} &= e^{i (\nu_p -\nu_3)t} \tilde{\rho}_{ca_2},
\nonumber\\
\rho_{a_1 b} &= e^{-i (\nu_2 +\nu_3)t} \tilde{\rho}_{a_1b},
\nonumber\\
\rho_{a_2 b} &= e^{-i \nu_2 t} \tilde{\rho}_{a_2b}, \nonumber \\
\rho_{a_1 a_2} &= e^{-i \nu_3 t} \tilde{\rho}_{a_1a_2}, \nonumber
\\ \rho_{b c} &= e^{i (\nu_1-\nu_p) t} \tilde{\rho}_{bc},
\nonumber \\ \rho_{a_1c} &= e^{-i \nu_p t} \tilde{\rho}_{a_1c}.
\end{align}

The equations of motion for the density matrix elements in the rotated frame
take the following form:
\begin{align}
\dot{\tilde{\rho}}_{a_1a_2} &= -\Bigl[i\,(\omega_{a_1a_2}- \nu_{3} ) +\half (\gamma_{1}+\gamma_{2})\Bigr]\tilde{\rho}_{a_1a_2} 
\nonumber \\
&\quad
- \frac{i}{2} \Omega_3\, e^{-i \varphi} e^{ i k x \cos \theta_3} (\tilde{\rho}_{a_1a_1} - \tilde{\rho}_{a_2a_2}) 
\nonumber \\
&\quad+ \frac{i}{2} \Omega_1  {\sin \kappa x}\,\tilde{\rho}_{ba_2} \nonumber \\ & \quad - \frac{i}{2} \Omega_2 e^{- i k x \cos \theta_2} \tilde{\rho}_{a_1b} 
+ \frac{i \mathcal{E}_p \wp_{a_1c}}{2\hbar} \tilde{\rho}_{ca_2}, 
\nonumber \\ 
\dot{\tilde{\rho}}_{a_1b} &= -\Bigl[i \, (\omega_{a_1b}- \nu_{1})+\half \gamma_{1}\Bigr]\tilde{\rho}_{a_1b} 
\nonumber \\
&\quad
- \frac{i}{2} \Omega_1 \sin{\kappa x}\, (\tilde{\rho}_{a_1a_1} - \tilde{\rho}_{bb}) 
\nonumber \\
&\quad- \frac{i}{2}\Omega_2 e^{i k x \cos \theta_2} \tilde{\rho}_{a_1a_2} 
+ \frac{i}{2}\Omega_3\, e^{- i \varphi} e^{ i k x \cos\theta_3} \tilde{\rho}_{a_2b} 
\nonumber \\ 
& \quad + \frac{i \mathcal{E}_p \wp_{a_1c}}{2\hbar} \tilde{\rho}_{cb}, 
\nonumber \\
\dot{\tilde{\rho}}_{a_2b} &= -\Bigl[i\,(\omega_{a_2b}- \nu_{2})+\half\gamma_{2}\Bigr]\tilde{\rho}_{a_2b} 
\nonumber \\
&\quad
- \frac{i}{2} \Omega_2\, e^{ i k x \cos \theta_2 } (\tilde{\rho}_{a_2a_2} - \tilde{\rho}_{b b}) 
\nonumber \\
&\quad- \frac{i}{2} \Omega_1 \sin{\kappa x}\,\tilde{\rho}_{a_2a_1} 
+  \frac{i}{2} \Omega_3\, e^{i \varphi}  e^{-i k x \cos \theta_3} \tilde{\rho}_{a_1b}, 
\nonumber \\
\dot{\tilde{\rho}}_{a_1c} &= -\Bigl[i\,( \omega_{a_1c}-\nu_{p}) + \half \gamma_1\Bigr] \tilde{\rho}_{a_1c} 
\nonumber \\
&\quad
+ \frac{i}{2} \Omega_3\, e^{- i \varphi} e^{ i k x \cos \theta_3} \tilde{\rho}_{a_2c} 
\nonumber \\
&\quad+ \frac{i}{2} \Omega_1 \sin \kappa x \tilde{\rho}_{bc} 
- \frac{i \mathcal{E}_p \wp_{a_1c}}{2\hbar} (\tilde{\rho}_{a_1a_1}-\tilde{\rho}_{cc}),
\nonumber \\ 
\dot{\rho}_{a_2c} &= -\Bigl[i [ \omega_{a_2c}-(\nu_{p}-\nu_{3})] + \half \gamma_2 \Bigr]\tilde{\rho}_{a_2c} 
\nonumber \\
&\quad+ \frac{i}{2} \Omega_3 e^{i \varphi}  e^{- i k x \cos \theta_3} \tilde{\rho}_{a_1c} 
+ \frac{i}{2}\Omega_2 e^{ i k x \cos \theta_2} \tilde{\rho}_{bc} 
\nonumber \\
&\qquad\qquad- \frac{i \mathcal{E}_p \wp_{a_1c}}{2\hbar} \tilde{\rho}_{a_2a_1}, 
\nonumber \\
\dot{\tilde{\rho}}_{bc} &= -\Bigl[i [ \omega_{bc}+(\nu_{1}-\nu_{p})] + \gamma_{bc}\Bigr] \tilde{\rho}_{bc} 
\nonumber \\
&\quad
+ \frac{i}{2} \Omega_1 \sin{ \kappa x}\, \tilde{\rho}_{a_1c} 
\nonumber \\
&\quad+ \frac{i}{2}\Omega_2 e^{- i k x \cos \theta_2} \tilde{\rho}_{a_2c} 
- \frac{i \mathcal{E}_p \wp_{a_1c}}{2\hbar} \tilde{\rho}_{ba_1}\,.
\label{Eq:completeset}
\end{align}
Now we use the definition of the detuning from Eq.~(\ref{Eq:Delta}) into Eq.~(\ref{Eq:completeset}) and after linearization with respect to the probe field we obtain Eq.~(\ref{eq:tilderhodot2}) which is only a set of equations useful for determining the susceptibility of the medium at the probe frequency.

Now we consider the details of the solution of Eq.~(\ref{eq:tilderhodot2}). 
This set of equations can be solved by writing them in the matrix form
\begin{equation}
\dot{R(t)} = - M R(t) + C\,,
\end{equation}
where $R(t)$ and $C$ are the column vectors and  $M$ is a matrix:
\begin{widetext}
\begin{align}
R &= \( \begin{matrix} \tilde{\rho}_{a_1c} & \tilde{\rho}_{a_2c} &
\tilde{\rho}_{bc}
\end{matrix}\)^{\rm T}\, , \nonumber \\
C &= \( \begin{matrix} i \frac{\mathcal{E}_p\wp_{a_1c}}{2\hbar} &
0 & 0
\end{matrix}\)^{\rm T} \, , \nonumber \\
M &= \nonumber \\
&\!\!\!\!\!\!\!\! \(\begin{matrix} \bigl(i \Delta + \half \gamma_1\bigr) 
& -\dfrac{i}{2}\Omega_3 e^{-i \varphi}  e^{i k x \cos \theta_3} 
& -\dfrac{i}{2} \Omega_1 \sin \kappa x \\ 
& & \\
-\dfrac{i}{2}\Omega_3 e^{i \varphi}  e^{-i k x \cos \theta_3} 
& \bigl(i \Delta + \half\gamma_2\bigr) 
& - \dfrac{i}{2}\Omega_2  e^{i k x \cos \theta_2} \\ 
& & \\
- \dfrac{i}{2} \Omega_1  \sin{\kappa x} &
-\dfrac{i}{2}\Omega_2  e^{- i k x \cos \theta_2} & i \Delta
\end{matrix}\).
\end{align}
The formal solution of such equations is given by
\begin{equation}
R(t) = \int_{-\infty}^{t} e^{- M (t-t') } C {\rm d}t' = M^{-1}
C. \label{eq:solution}
\end{equation}
Finally by using Eq.~(\ref{eq:solution}), we obtain the solution, Eq.~(\ref{eq:tilderhoa1c}).
\end{widetext}


\begin{thebibliography}{99}

\bibitem{welch9193} K. D. Stokes, C. Schnurr, J. R. Gardner, M. Marable,
G. R. Welch, and J. E. Thomas, Phys. Rev. Lett. {\bf 67}, 1997 (1991);
J. R. Gardner, M. L. Marable, G. R. Welch, and J. E. Thomas, Phys.
Rev. Lett. {\bf 70}, 3404 (1993).

\bibitem{ChuMetcalf} S. Chu and C. Wieman, J. Opt. Soc. Am. B {\bf 6}, 2020 (1989);
H. Metcalf and P. Van der Straten, Phys. Rep. {\bf 244}, 204 (1994).

\bibitem{Collins96} G. P Collins, Phys. Today {\bf 49} (3), 18 (1996).

\bibitem{lithography} M. O. Scully and K. Dr\"{u}hl Phys. Rev. A {\bf 25}, 2208 (1982); 
U. W. Rathe and M. O. Scully Lett. Math. Phys. {\bf 34}, 297 (1995); A. N. Boto, P. Kok, D. S. Abrams, S. L. Braunstein, C. P. Williams,
and J. P. Dowling  Phys. Rev. Lett. {\bf 85}, 2733 (2000); G. S. Agarwal and M. O. Scully,   Opt. Lett. {\bf 28}, 462 (2003).

\bibitem{KapaleWavefunction} K. T. Kapale, S. Qamar, and M. S. Zubairy, Phys Rev. A {\bf 67},
023805 (2003).

\bibitem{WallsZollerWilkens} P. Storey, M. Collett, and D. Walls, Phys. Rev.
Lett. {\bf 68}, 472 (1992); Phys. Rev. A {\bf 47}, 405 (1993); M. A. M. Marte
and P. Zoller, Appl. Phys. B {\bf 54}, 477 (1992); 
S. Kunze, G. Rempe, and M. Wilkens, Europhys. Lett.
{\bf 27}, 115 (1994).


\bibitem{Walls95} R. Quadt, M. Collett, and D. F. Walls, Phys. Rev. Lett. 
{\bf 74}, 351 (1995).


\bibitem{Kunze97} S. Kunze, K. Dieckmann, and G. Rempe, Phys. Rev. Lett. 
{\bf 78}, 2038 (1997).

\bibitem{Kien97} F. Le Kien, G. Rempe, W. P. Schleich, and M. S. Zubairy,
Phys. Rev. A {\bf 56}, 2972 (1997).


\bibitem{Thomas90} J. E. Thomas, Phys. Rev. A {\bf 42}, 5652 (1990).

\bibitem{Bigelow97} P. Rudy, R. Ejnisman, and N. P. Bigelow, Phys. Rev. Lett 
{\bf 78}, 4906 (1997).


\bibitem{Zoller96} M. Holland, S. Marksteiner, P. Marte, and P. Zoller, Phys. Rev.
Lett. {\bf 76}, 3683 (1996).

\bibitem{Herkomer97} A. M. Herkommer, W. P. Schleich, and M. S. Zubairy, J. Mod.
Opt. {\bf 44}, 2507 (1997).


\bibitem{QamarPRA2000} S. Qamar, S.-Y. Zhu, and M. S. Zubairy, Phys. Rev. A
{\bf 61}, 063806 (2000).


\bibitem{QamarOC2000} S. Qamar, S.-Y. Zhu, and M. S. Zubairy, Opt.
Commun. {\bf 176}, 409 (2000).


\bibitem{CohContrlSE} E. Paspalakis, C. H. Keitel, and P. L. Knight, Phys. Rev.
A {\bf 58}, 4868 (1998); S.-Y. Zhu and M. O. Scully, Phys. Lett.
A {\bf 201}, 85 (1995); E. Paspalakis, and P. L. Knight, Phys. Rev. Lett. {\bf 81},
293 (1998); H. Lee, P. Polynkin, M. O. Scully, and S.-Y. Zhu,
Phys. Rev. A {\bf 55}, 4454 (1996); S.-Y. Zhu, Quantum Opt. {\bf 7}, 385 (1995);
K. T. Kapale, M. O. Scully, S.-Y. Zhu, and M. S. Zubairy, Phys. Rev. A {\bf  67},  023804 (2003).  


\bibitem{GhafoorPRA2000} F. Ghafoor, S.-Y. Zhu, and M. S. Zubairy, Phys. Rev. A
{\bf 62}, 013811 (2000).


\bibitem{GhafoorPRA2002} F. Ghafoor, S. Qamar, and M. S. Zubairy, Phys. Rev. A {\bf 65}, 
043819 (2002).


\bibitem{Knight2001} E. Paspalakis, and P. L. Knight, Phys. Rev. A {\bf 63}, 
065802 (2001).


\bibitem{EIT} K. J. Boller, A. Imamoglu, and S. E. Harris, Phys. Rev. Lett.
{\bf 66}, 2593 (1991); J. E. Field, K. H. Hahn, and S. E. Harris, Phys. Rev.
Lett. {\bf 67}, 3062 (1991).

\bibitem{SahraiGroupVel} M. Sahrai, H. Tajalli, K. T. Kapale, and M. S. Zubairy,
Phys. Rev. A {\bf 70}, 023813 (2004).


\bibitem{Meystre} P. Meystre, and M. Sargent III, {\it  Elements of Quantum Optics},
3rd ed. (Springer-Verlag, Berlin, 1999).


\end{thebibliography}
\end{document}